\documentclass[ %
 reprint,
 amsmath,amssymb,showkeys,
 aps, superscriptaddress,
]{revtex4-2}

\usepackage{graphicx}
\usepackage{dcolumn}
\usepackage{bm}
\usepackage{braket}

\begin{document}

\title{Indirect Indication of the KCBS-Type Quantum Contextuality by the CHSH-Bell Test}

\author{Fırat Diker}
 \email{firatdiker@alumni.sabanciuniv.edu}
\affiliation{Faculty of Engineering and Natural Sciences, Sabanci University, Istanbul 34956, Turkey }
\affiliation{International Centre for Theory of Quantum Technologies, University of Gdansk, Gdansk 80-309, Poland }

\author{Antonio Mandarino}
\affiliation{International Centre for Theory of Quantum Technologies, University of Gdansk, Gdansk 80-309, Poland }

\date{\today}

\begin{abstract}
In this work, we show that the well-known Bell inequality in the Clauser-Horne-Shimony-Holt (CHSH) formulation serves not only as a non-locality test but also as a tool to prove KCBS-type contextuality. For this purpose, we investigate the symmetric family of two-qubit states that can be mapped to qutrit states (three-level quantum states) and show that they exhibit a contextual description, as evidenced by the KCBS-type inequality. Later, we apply the CHSH test to them and find a new non-contextuality bound for this test. This shows that the CHSH inequality can be modified to serve as a KCBS-type contextuality test by changing the limit. Also, the number of measurements required is four, fewer than the KCBS test's five. We discuss the new bound concerning the monogamy problem.
\end{abstract}


\maketitle

\section{\label{sec1} INTRODUCTION}

Quantum non-locality and contextuality are fundamentally important properties unique to Quantum Theory (QT) and have been a topic of interest in the scientific community since the pioneering works \cite{Gleason,1_Bell,2,3}. A solid understanding of these fundamental concepts and their relations is essential in QT. These characteristics are counterintuitive and incompatible with the classical approach because, in quantum physics, the measurement context does affect the outcome. Also, a quantum measurement is independent of local restrictions imposed by classical physics. These quantum-unique features are intriguing; consequently, new questions have arisen about whether QT is a complete theory. Therefore, it has been argued that there may be some missing pieces of the puzzle, namely, the hidden variables; however, this argument has later been denied \cite{1_Bell,2,3,4_Bell,5,6}. The most basic examples of classical inequalities regarding the number of measurements are the Clauser-Horne-Shimony-Holt (CHSH) inequality and the Klyachko-Can-Binicio\u{g}lu-Shumovsky (KCBS) inequality \cite{CHSH,KCBS1,KCBS4}. These measurement scenarios allow us to test quantum correlations (contextuality and non-locality). We will mention the CHSH and the KCBS measurement scenarios and show the relationship between the non-locality and contextuality of the symmetric two-qubit states. It is known that some qutrit states exhibit contextuality intrinsically, and one can say that they are compatible with quantum predictions \cite{8,9}. Since they are single-particle states, entanglement is not involved in this case. Increasing the number of measurements lets us observe contextuality for every qutrit state \cite{10,11}.

The outline of this paper is as follows: We will briefly discuss the KCBS inequality. Later, we will introduce the family of symmetric two-qubit states and the corresponding qutrit states for each. After that, we will mention the maximal violation of the KCBS and CHSH inequalities for a given concurrence (degree of entanglement). Using the concurrence-based correlation, we will finally derive a new non-contextuality bound for the CHSH test. 

Bell formulated the locality principle in terms of testable inequalities and demonstrated that two-qubit quantum systems generically violate them, exhibiting non-locality \cite{1_Bell}. The most widely studied and experimentally implemented instance is the Bell-CHSH inequality \cite{CHSH}, which involves four dichotomic measurements on a bipartite two-qubit system. A landmark result due to Gisin \cite{17} established that every entangled pure two-qubit state violates the Bell-CHSH inequality. This connection was made quantitative by Wotters \cite{conc-bell1} and Verstraete and Wolf  \cite{conc-bell2}, who showed that the maximum Bell-CHSH violation is an increasing function of the concurrence $C$, specifically $\beta_{\max} = 2\sqrt{1+C^2}$. This result firmly links entanglement—as measured by concurrence or equivalently the entanglement of formation \cite{conc-bell1} to the degree of Bell-CHSH non-locality: greater entanglement directly implies a larger violation of the classical bound. However, this correspondence is not universal across all Bell inequalities; certain entangled states need tailored Bell-type inequalities, or the maximally entangled states do not violate certain inequalities \cite{not-violated1,not-violated2,not-violated3,not-violated4}, revealing that entanglement is necessary but not always sufficient for Bell non-locality. 

Research into higher-dimensional systems has further enriched this picture. For bipartite $d \times d$ systems (qudits), generalizations of the Bell-CHSH inequality reveal that higher-dimensional entanglement can yield larger violations of locality bounds \cite{not-violated1}, and the entanglement of formation continues to serve as the central entanglement monotone governing the strength of quantum correlations \cite{Terhal_2000}. In dimensions $d \geq 3$, however, a qualitatively new phenomenon emerges: even a single quantum system can exhibit non-classical correlations that have no bipartite entanglement interpretation. This was established by Kochen and Specker \cite{2,3}, who proved that in Hilbert spaces of dimension $d \geq 3$, the outcomes of quantum measurements cannot be assigned definite values independent of the measurement context. This property—contextuality—is a feature of the quantum state space itself and does not require spatial separation between subsystems. It is therefore a strictly more general form of non-classicality than Bell non-locality, and understanding its relationship to entanglement and to Bell-CHSH violations is central to characterizing the full landscape of quantum correlations.

Recent work has further clarified the geometry of the CHSH local set and its relationship with other forms of non-classicality. In particular, the authors in \cite{Gigena_2023} provided a constructive and geometrically intuitive description of the local polytope and its facets in the standard bipartite Bell scenario with two dichotomic measurements per party, which is useful for framing any modified CHSH threshold in terms of the underlying classical region. In parallel, in \cite{Scala_2024}, random pure two-qubit systems were examined, and it was shown that the violation of a contextuality inequality is strongly reflected in Bell non-classicality, reinforcing the idea that contextual and nonlocal signatures are often intertwined in experimentally relevant two-qubit settings. More recently, in \cite{Mandarino_2023}, the authors studied the robustness of the CHSH-Bell violation to filtering and state-preparation imperfections for random mixed states, emphasizing how the observed non-locality depends not only on the abstract state but also on the effective experimental accessibility of the chosen setup. Taken together, these results support the present analysis by showing that the CHSH scenario can be used not only as a witness of Bell non-locality, but also as a refined diagnostic tool for distinguishing different layers of quantumness, especially when one exploits symmetry-restricted two-qubit states mapped onto effective qutrit descriptions.

In this work, we show that the CHSH test can be employed as an indirect witness of KCBS-type contextuality for symmetric two-qubit states admitting an effective qutrit description. Combining the concurrence-dependent KCBS and CHSH bounds, we obtain a new CHSH-based non-contextuality limit, $\beta \leq \sqrt{24/5}$, which identifies a regime of states that are nonlocal yet noncontextual. This provides a sharper operational distinction between classicality, Bell non-locality, and contextuality. The remainder of the paper is organized as follows: Sec.\ref{sec2} presents the KCBS and CHSH scenarios and derives the new bound; Sec.\ref{sec3} analyzes the monogamy question in the single-qutrit and two-qubit settings; and Sec.\ref{sec4} discusses the physical significance of the result and concludes the paper.

\begin{figure}[t]
\centering
\includegraphics[width=\columnwidth]{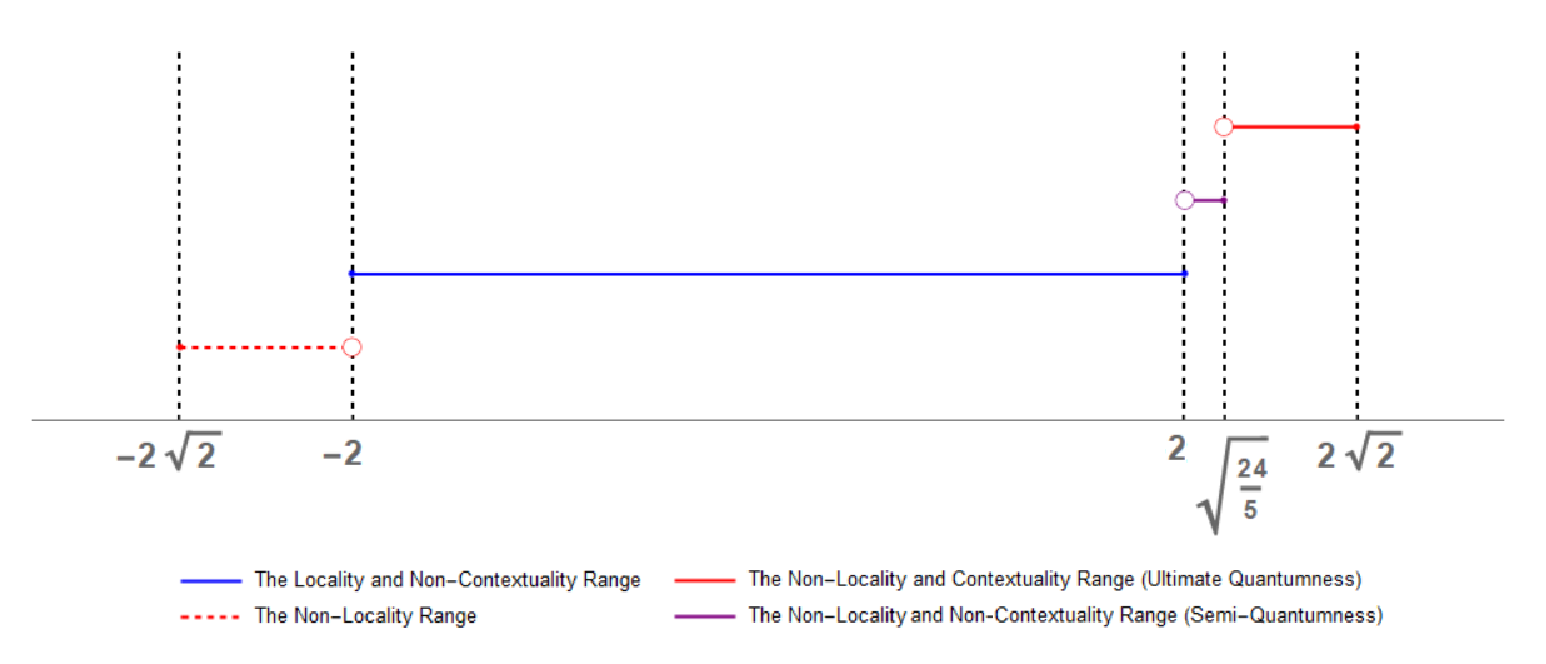}
\caption{(color online) These are the measurement ranges for the CHSH operator, with dashed vertical black lines separating them. The separation is based on their (non-)quantumness; in other words, it depends on whether the measurements yield results inside or outside the classical range. The dashed red line shows the non-locality range; the blue solid line is for the locality and non-contextuality region; the purple solid line is for the measurements yielding non-local and non-contextual results; and finally, the red solid line shows us the non-locality and contextuality region.}
\label{fig1}
\end{figure}

\section{THE CHSH AND KCBS INEQUALITIES}

The CHSH and KCBS inequalities are two of the most standard tools for revealing nonclassical correlations in quantum theory, but they probe conceptually different regimes. The CHSH inequality is a Bell inequality for bipartite systems with dichotomic measurements, while the KCBS inequality is a contextuality inequality for a single qutrit with five compatible observables arranged in a cyclic compatibility structure. Although both inequalities separate quantum predictions from classical hidden-variable models, they refer to different notions of classicality: CHSH tests local realism, whereas KCBS tests noncontextuality. For this reason, their comparison is especially useful when one studies systems that can be viewed either as effective qutrits or as symmetric two-qubit states.

In the CHSH scenario, two distant parties each perform one of two possible binary measurements, usually denoted by $A_1, A_2$ for the first party and $B_1, B_2$ for the second. The CHSH expression is
\begin{equation}
\beta = \langle A_1 B_1 \rangle + \langle A_1 B_2 \rangle + \langle A_2 B_1 \rangle - \langle A_2 B_2 \rangle,
\end{equation}
up to equivalent relabelings of the observables. Any local hidden-variable model satisfies the bound
\begin{equation}
|\beta| \leq 2,
\end{equation}
whereas quantum theory allows larger values, with the Tsirelson bound $|\beta| \leq 2\sqrt{2}$. Thus, a violation of the CHSH inequality certifies Bell nonlocality. For pure two-qubit states, the amount of CHSH violation is directly linked to the concurrence, so that stronger entanglement typically produces larger nonlocal correlations.

The KCBS inequality concerns five observables $A_1,\dots,A_5$ with outcomes $\pm 1$, such that each observable is compatible with its nearest neighbors in the cycle. A standard form of the inequality is
\begin{equation}
\langle A_1 A_2 \rangle + \langle A_2 A_3 \rangle + \langle A_3 A_4 \rangle + \langle A_4 A_5 \rangle + \langle A_5 A_1 \rangle \geq -3.
\end{equation}
This bound holds for any noncontextual hidden-variable model. Quantum mechanics violates it for suitable qutrit states and measurement choices, reaching a minimum value close to $5-4\sqrt{5}$ in the standard KCBS setting. Unlike CHSH, the KCBS inequality does not require spatial separation or entanglement; its violation reflects the impossibility of assigning preexisting context-independent values to all observables simultaneously.

The relevance of these two inequalities in the present work is that they can be connected through symmetric two-qubit states that admit an effective qutrit description. In such cases, the concurrence acts as the common parameter controlling both Bell nonlocality and KCBS-type contextuality. This shared dependence allows translating the classical CHSH boundary into a new threshold for contextuality, thereby identifying regimes in which a state is nonlocal but not contextual, as well as those in which the two quantum features coexist. This relation is central to the discussion in the following sections.

\section{\label{sec2} THE CORRELATION BETWEEN THE OUTCOMES OF THE KCBS AND CHSH TESTS THROUGH CONCURRENCE}

The simplest example of a non-contextuality test is the KCBS scenario, which includes five spin-1 measurements in total \cite{KCBS1,KCBS4}. Firstly, it was presented in \cite{KCBS1} and then explained comprehensively in \cite{KCBS4}. This is a state-dependent scenario, and the KCBS inequality is given as
\begin{equation}
\langle A_1 A_2 \rangle + \langle A_2 A_3 \rangle + \langle A_3 A_4 \rangle + \langle A_4 A_5 \rangle + \langle A_5 A_1 \rangle \geq -3
\label{eq1}
\end{equation}
where $A_i = 2 S_i^2 - 1$ ($S_i$ are the spin-$1$ operators). In real space ($\mathbb{E}^3$), there are five directions along which spin measurements can be performed, and $A_i$ and $A_{i+1}$ are co-measurable due to orthogonality relations. The vectors taken along these directions form a pentagram. Each term is the average value of two measurements ($A_i$ and $A_{i+1}$) performed together. We obtain results exceeding the lower limit; in other words, contextuality is a quantum-unique property of qutrit states. Maximal contextuality of the KCBS scenario is observed when the neutrally polarized spin state $\vert 0 \rangle$ is tested. One finds the quantum lower limit close to $-4$ ($\cong -3,94$).

In the KCBS paper, the authors have mentioned the possibility of symmetric two-qubit usage for testing its KCBS-type contextuality \cite{KCBS4}. They have even given the concurrence (the degree of entanglement) for these states to violate the KCBS inequality \cite{KCBS4,conc-bell1}. For $C > 0,447$, symmetric two-qubit states violate the KCBS inequality, exhibiting contextuality. Later, the degree of KCBS-type contextuality has been found for a known concurrence \cite{diker2022degree}. In this work, one uses symmetric two-qubit states corresponding to effective qutrit states, and measures their spins in the KCBS scenario. The definition of a symmetric two-qubit state is the following:
\begin{equation}
    \ket{\psi} = a \ket{00} + \frac{b}{\sqrt{2}} \big(\ket{01} + \ket{10} \big) + c \ket{11}
\end{equation}
where $a$, $b$, and $c$ are complex probability amplitudes, satisfying the normalization condition (${|a|}^2 + {|b|}^2 + {|c|}^2 = 1$). The state $\ket{\psi}$ corresponds to the general definition of a qutrit state ($\ket{\psi^*} = a \ket{1} + b \ket{0} + c \ket{-1}$). By taking into account all possible qutrit states, one obtains the degree of contextuality in terms of concurrence as follows \cite{diker2022degree}:
\begin{equation}
S^{min}_{KCBS} = \left(5-3 \sqrt{5}\right) C - \sqrt{5}.
\label{SminC}
\end{equation}
One obtains the possible minimum value when applying the KCBS test for a given concurrence. Equation \ref{SminC} is a simple linear equation that gives us the degree of contextuality for a qutrit state with an embedded entanglement. Concurrence can take values between 0 and 1 ($0 \leq C \leq 1$), and we may check for $C = 1$ and $C = 0$, i.e., the maximally entangled and non-entangled quantum states, respectively. When $C=0$, $S^{min}_{KCBS} = - \sqrt{5}$. That is the limit value for non-entangled states. Recall the non-contextuality limit ($=-3$), and the new bound given for local states is higher than this. We observe non-contextual and non-local states between these limits, i.e., semi-quantumness. For maximal entanglement ($C=1$), $S^{min}_{KCBS} = 5-4 \sqrt{5} $, which corresponds to the maximal contextuality in the KCBS scenario. We find $C \cong 0.447$ for the non-contextuality limit (-3), and both of these concurrence values have already been given in the KCBS paper \cite{KCBS4}. The degree of entanglement is high for us to observe quantum contextuality.

Similarly, for a given concurrence, the maximal non-locality in the CHSH measurement scenario is given by the following: 
\begin{equation}
 \beta = 2 \sqrt{1+C^2}, 
 \label{beta}
\end{equation} 
and $\beta$ is the expectation value of the CHSH operator \cite{conc-bell1,conc-bell2}. Equations \ref{SminC} and \ref{beta} are correlated through concurrence, i.e., entanglement is related to both quantum contextuality and non-locality. This leads us to ask whether one can test the KCBS-type quantum contextuality in the CHSH test.

\begin{figure}[b]
\includegraphics[width=1.0\columnwidth]{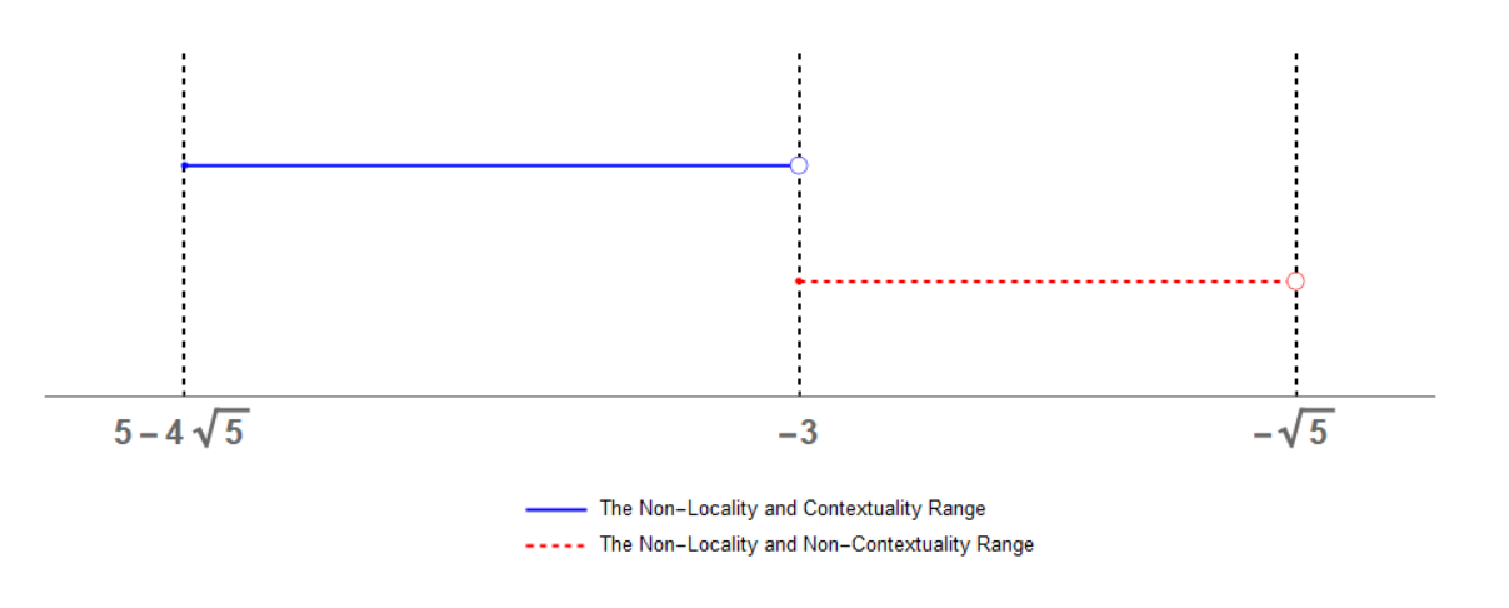}
\caption{(color online) In this graph, we present the ranges of measurement outcomes of the KCBS operator, and use dashed vertical black lines to separate them. The separation is based on their (non-)classicality; in other words, it depends on whether the measurements yield results of quantum correlations. The dashed red line shows the non-locality and non-contextuality range; the blue solid line is for the non-locality and contextuality region. The KCBS operator allows us to observe the boundary ($S=-\sqrt{5}$) between the CHSH-type non-locality and local correlations. $S=-3$ is the usual non-contextuality bound for the KCBS measurements; $S= 5 - 4 \sqrt{5}$ is the quantum lower limit, corresponding to the maximal quantum contextuality.}
\label{fig2}
\end{figure}

We need to find the concurrence value for a symmetric two-qubit state to give the classical limit for the KCBS inequality. One has to solve Equation \ref{SminC} for $S=-3$, which gives $C \cong 0.447$. We plug this value in Equation \ref{beta} and obtain  $\beta = \sqrt{\frac{24}{5}}$ ($\cong 2.191$). This gives us the CHSH-type non-contextuality bound, and one may write
\begin{equation}
    \beta \leq \sqrt{\frac{24}{5}},
\end{equation}
which is the CHSH-type non-contextuality inequality. So, when $\sqrt{\frac{24}{5}} < \beta \leq 2 \sqrt{2}$, we observe contextual and non-local states; when $2 < \beta \leq \sqrt{\frac{24}{5}}$, quantum states are non-local and non-contextual; and finally, when $\beta \leq 2$, we obtain local and non-contextual states when tested (no quantum correlation). Our result imposes a stricter separation between quantumness and classicality. One can easily see the range within which we may observe non-locality together with non-contextuality. In other words, the states exhibit semi-quantumness. So, violating the CHSH inequality does not necessarily mean non-classicality in all aspects. The boundaries, including the new non-contextuality limit, are illustrated in Fig. \ref{fig1}. Also, the locality bound of the CHSH inequality can be indirectly seen in the KCBS test (Fig. \ref{fig2}). 

Freedman and Clauser have demonstrated the CHSH test \cite{PhysRevLett.28.938}. Later, Aspect showed the violation of a Bell inequality with high accuracy \cite{PhysRevLett.49.91}, as compared with the demonstration of Freedman and Clauser \cite{PhysRevLett.28.938}. In another work, the observers were independent of each other, since Aspect used randomly changing polarization settings while photons were flying \cite{5}. Aspect proposed an improved technique to switch the photons into two apparatus branches on timescales short enough \cite{PhysRevD.14.1944}. In Zeilinger's experiment, they set strict local conditions and tested the inequality using improved techniques \cite{PhysRevLett.81.5039}, with the distance between observers at 400 m. In 2006, using an optical free-space link, the Zeilinger group established a secure key between two islands separated by 144 km \cite{ursin2007entanglement}. Polarization-entangled photon pairs were used to test the CHSH inequality, demonstrating a violation of locality. The contextuality test in the CHSH scenario is realizable since the CHSH measurements have already been performed in the previous experiments.

\section{\label{sec3} IS THIS A MONOGAMY PROBLEM?}

Having established the concurrence-dependent threshold, we next examine whether the relation between KCBS-type contextuality and CHSH-type non-locality can be interpreted as a monogamy constraint. We distinguish between the effective single-qutrit picture and the genuine two-qubit scenario, since the physical meaning of the correlations differs in these two cases.

\subsection{A Single Qutrit Case}

As mentioned before, for a symmetric two-qubit state to violate the KCBS inequality, its concurrence must be above a certain value. As long as concurrence is non-zero, it is evident from Equation \ref{beta} that a quantum state violates the CHSH inequality and exhibits non-locality. One may define qutrit states as imaginary two-qubit states, provided there is symmetry between the qubits. So, one may calculate concurrence for these qubits and obtain the degree of entanglement. Indeed, there is no separation as in the case of real two-qubit states, and one cannot talk about entanglement between particles since there is only one particle, a qutrit. This is more likely an imaginary entanglement, called self-entanglement or embedded entanglement \cite{Can_2005}. Entanglement implies non-locality in the CHSH measurement scenario \cite{17}, and one may naturally ask whether this is a monogamy problem between non-locality and contextuality for qutrit states. The answer is 'no' because there is no entanglement here; i.e., we cannot talk about non-local correlations. When the degree of entanglement (concurrence) exceeds a certain threshold, the KCBS and CHSH inequalities are violated simultaneously; however, even if we assume we can simulate the CHSH test using spin-1 operators and apply both the KCBS and CHSH tests, there is only a single qutrit. To investigate possible monogamy among quantum correlations, we need to examine the relationships between the degrees of non-locality and contextuality. Here, we observe only quantum contextuality for a single qutrit, indicating the absence of non-local correlations. 

The drawback of our approach is that we lack the spin-1 measurement operators corresponding to the spin-1/2 operators of the CHSH test. We already know that the usual spin-$1/2$ operators are used in the CHSH scenario, whereas the spin-1 measurements are performed in the KCBS scenario. So, one needs to find the corresponding CHSH operator constituted by spin-1 operators.

\begin{figure}[t]
\includegraphics[width=.7\columnwidth]{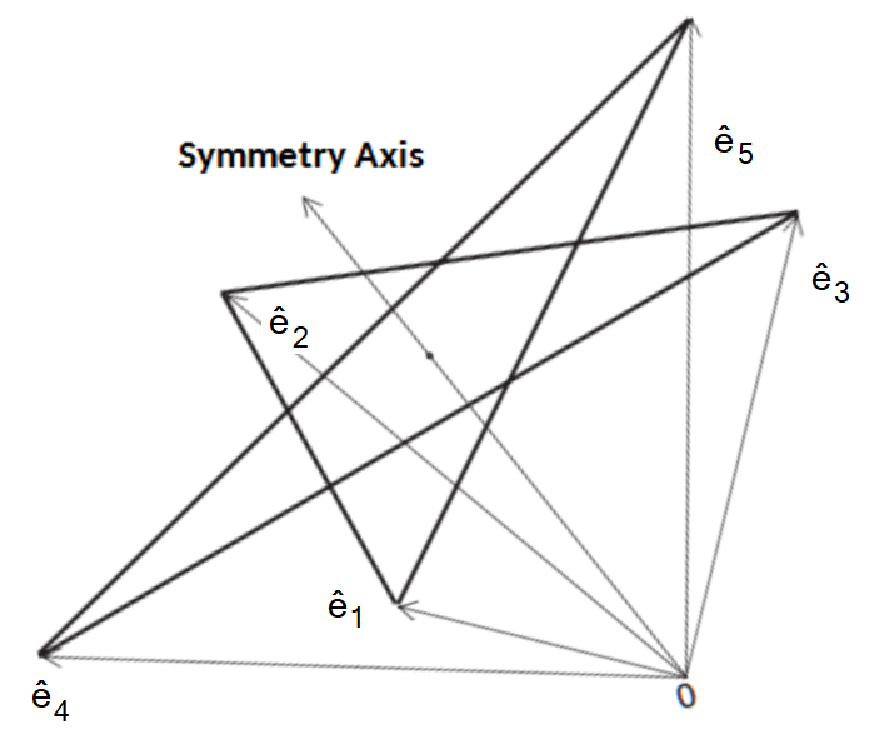}
\caption{(color online) The pentagram shows the compatibility relations among the spin operators in the KCBS scenario. Each $\hat{e}_i$ represents the direction along which spin-1 measurements are performed. Each vertex corresponds to a KCBS spin operator, and each line indicates the compatibility between the two operators.}
\label{fig3}
\end{figure}

\subsection{A Two-Qubit Case}

In the previous subsection, we discussed our result, accounting for the monogamy problem for a single qutrit. We do not have the real spin-1 operators corresponding to the CHSH operators. From self(embedded)-entanglement, we deduce that CHSH- and KCBS-type contextuality can be observed simultaneously for qutrits.

In this subsection, we investigate the case of real two-particle states that were previously treated as imaginary and corresponded to real qutrits in the previous case. As long as these two-qubit states are symmetric, they correspond to effective qutrit states. Here, we interchange the roles of two-qubit states and qutrits. So, in this case, the CHSH-type non-locality and the KCBS-type contextuality can be observed simultaneously, and the relation between them is no longer intrinsic since we have a bipartite state. Nevertheless, we cannot say that there is a polygamous relation between CHSH-type non-locality and KCBS-type contextuality, because both are non-local and contextual.

Recall that two-qubit states must be symmetric to correspond to effective qutrit states. As long as two qubits are entangled, they violate the CHSH locality inequality, exhibiting non-locality. For $C > 0.447$, they also violate the KCBS non-contextuality inequality; so above this concurrence limit, we observe both the CHSH-type non-locality and the KCBS-type contextuality. We also have the necessary real spin-1/2 operators that can be mapped onto the 3-dimensional Hilbert space to define effective spin-1 operators. All we need to do is perform spin-1/2 measurements on qubits along the directions forming the pentagram of the KCBS scenario (Fig. \ref{fig2}).

\section{\label{sec4} CONCLUSION}

The monogamy problem has been examined \cite{PhysRevLett.112.100401,PhysRevLett.130.040201}, and the monogamy relation between non-locality and contextuality does not always hold \cite{PhysRevLett.130.040201}. Both non-local and contextual correlations can be observed simultaneously. Our result for two-qubit states does not say anything about monogamy; however, we have observed that two-qubit states can exhibit both KCBS-type and CHSH-type non-local correlations. These correlations can also be called contextual. One can easily use real two-qubit states to run both the KCBS and CHSH tests and observe both CHSH- and KCBS-type non-local (contextual) correlations. One does not need to perform KCBS measurements, since the earlier non-contextuality limit explicitly shows the KCBS-type non-locality of bipartite states.

In conclusion, the correlation between the KCBS-type contextuality and the CHSH-type non-locality via entanglement allows us to reduce the number of measurements from five to four to observe the KCBS-type contextuality, and one performs spin-1/2 measurements on symmetric two-qubit states. The CHSH test has been well studied \cite{PhysRevLett.28.938, PhysRevLett.49.91, 5, PhysRevD.14.1944, PhysRevLett.81.5039, ursin2007entanglement} since it was first published. However, we did not know it could reveal this kind of contextuality. Our result allows us to draw a more precise boundary between quantumness and classicality. The range within which one can observe non-locality and non-contextuality gives us the states exhibiting semi-quantumness. The classical limits of the CHSH and KCBS tests give us different concurrence values.
We have shown that the CHSH test can be reinterpreted, for symmetric two-qubit states, as an indirect probe of KCBS-type contextuality through the concurrence. By combining the known concurrence dependence of the KCBS and CHSH maxima, we derived the threshold $\beta \leq  \sqrt{24/5}$, which separates local-noncontextual states from non-local states that remain noncontextual. The result sharpens the distinction between Bell non-locality and contextuality and suggests a more economical experimental route to contextuality witnessing in two-qubit platforms.

\section{ACKNOWLEDGMENT}
This work was supported by the Scientific and Technological Research Council of Türkiye (TÜB\.{I}TAK) as part of International Postdoctoral Research Fellowship Program (Program No: 2219,
Reference No: 53325897-115.02-644009).
AM acknowledges the IRA Programme, project no.~FENG.02.01-IP.05-0006/23, financed by the FENG program 2021-2027, Priority FENG.02, Measure FENG.02.01., with the support of
the FNP.

\nocite{*}

\bibliography{apssamp}

\end{document}